# Enhanced laser-driven electron beam acceleration due to ionization-induced injection

Song Li, Nasr A. M. Hafz[*], Mohammed Mirzaie[***], Thomas Sokollik, Ming Zeng, Min Chen, Zhengming Sheng, and Jie Zhang[**]

*Key Laboratory for Laser Plasmas (Ministry of Education) and Department of Physics and Astronomy, Shanghai Jiao Tong University, Shanghai 200240, China*

Corresponding authors: [*]nasr@sjtu.edu.cn, and [**]jzhang1@sjtu.edu.cn

[***]Equivalent First Author

**Abstract:** We report an overall enhancement of a laser wakefield acceleration (LWFA) using the ionization injection in a mixture of 0.3% nitrogen gas in 99.7 % helium gas. Upon the interaction of 30-TW, 30-fs laser pulses with a gas jet of the above gas mixture, >300 MeV electron beams were generated at a helium plasma densities of $3.3$–$8.5 \times 10^{18}$ cm$^{-3}$. Compared with the electron self-injection in pure helium gas jet, the ionization injection has led to the generation of electron beams with higher energies, higher charge, lower density threshold for trapping, and a narrower energy spread without dark current (low energy electrons) or multiple bunches. It is foreseen that further optimization of such a scheme is expected to bring the electron beam energy-spread down to 1 %, making them suitable for driving ultra-compact free-electron lasers.

## References and links

## 1. Introduction

Attributed to the outstanding advances in ultra-short high-power laser technologies [1], laser wakefield acceleration (LWFA) [2] has been realized and is now under consideration as a basis for the next generation high energy electron-positron colliders [3,4]. Over the past decade [5–9], major breakthroughs have been experimentally achieved and led to significant enhancements in the quality of electron beams in terms of maximum energy, divergence angle, pointing stability, shot-to-shot reproducibility, and charge. A recent experiment have generated quasi-monoenergetic electron beams with an energy spectrum peaked at 2 GeV via the interaction of petawatt (PW) laser pulses with a gas cell of $10^{17}$-cm$^{-3}$ density [10].

In LWFA, the process of electron injection into the subtle accelerating structures is essential for the quality of resulting electron beams. Significant efforts have been devoted to control the electron injection into the wakewave. For example, the optical injection [11–13] pre-accelerates some electrons to the phase velocity (almost the speed of light *c*) of the wakefield, then these electrons can be further accelerated continuously in the accelerating and focusing phase; the density-ramp injection [14–18] or bubble evolution injection [19] can slow down the wake so that some electrons can easily catch up with the acceleration phase. Recently, the ionization-induced injection scheme was proposed [20–22]; it utilizes the large difference in ionization potentials between the outer and inner shell electrons of trace high-Z atoms (such as nitrogen or oxygen) mixed in low-Z atoms (usually helium or hydrogen) to control the initial phase of the electrons ionized from the inner shell. This scheme has the advantage of being experimentally simple and been experimentally demonstrated by several groups. In 2010, Clayton *et al.* [23] accelerated electrons up to 1.45 GeV continuous energy spectrum using a 1.3-cm long gas cell by mixing 3% $CO_2$ in helium at the plasma density below $1.5 \times 10^{18}$ cm$^{-3}$ under the matched and self-guiding conditions. In 2013, Mo *et al.* [24] obtained 0.5–1-GeV electron bunches (quasimonoenergetic peaks with long tails of background electrons or multiple bunches) from LWFA via ionization induced injection using mixtures of 4%–10% $CO_2$ in He. To enhance the beam quality, the ionization injection was demonstrated by the two-stage accelerating configuration, where the electron injection and acceleration were separated and manipulated in two different gas cells. Liu *et al.* [25] reported near-GeV electron beams using two gas cells filled with a mixture of 94% He and 6% $O_2$ (first) and pure He gas (second). Pollock *et al.* [26] generated a ~0.5 GeV electron beam through two-stage gas cell, where a mixture of 99.5% He and 0.5% $N_2$ was filled in the first 3-mm injection stage, and the second 5-mm acceleration stage contained pure He. As mentioned above, it is noticed that the ionization injection in the first-stage (which contains the gas mixture) in both [25, 26] as well as in the [23, 24] experiments could only generate low-quality electron beams due to the continuous injection of electrons ionized from the doped gas into the wake of the host one.

Here, we observed the generation single electron bunches having high-quality and dark-current free energy spectrum via the ionization injection in a few-mm long single-stage LWFA by adding *ultralow* amounts of nitrogen gas (0.3 %) into a helium (99.7 %) gas jet. Upon the interaction of 30-TW 30-fs laser pulses having unmatched spot sizes with the above gas mixture, > 300 MeV electron beams were generated at helium plasma densities of $3.3$–$8.5 \times 10^{18}$ cm$^{-3}$. Compared with the electron self-injection in pure helium gas jet, the ionization injection of electrons from the low concentration nitrogen has led to the generation of beams with significantly higher energies, higher charge, lower density threshold for trapping, and more importantly beams without dark currents (low energy electrons) or multiple bunches. Thus we are reporting an overall enhancement in the laser wakefield acceleration. It is obvious that an optimized single-stage scheme is easier to implement than the dual-stage or the optical injection schemes; we experimentally propose this simple method of generating high-quality beams, in support of our recently-published simulations in Ref [27]. It is foreseen that further optimization of such a scheme is expected to bring the electron beam energy-spread down to 1 %, making them suitable for driving ultra-compact free-electron lasers [28]. The experimental setup is given in section 2, results and discussions are presented in section 3, and section 4 is for the conclusions.

## 2. Experimental setup

The experiment was conducted using a Ti:sapphire terawatt laser system at the Key Laboratory for Laser Plasmas of Shanghai Jiao Tong University in China. The basic experimental arrangement has been introduced in an earlier publication Ref. [29], here we give a brief description on the enhancement in the setup. In the present experiment, 28–36 TW, 800 nm laser pulses were focused by a $f$/20 off-axis parabolic mirror onto a 4-mm supersonic gas jet which can blowout a prepared mixture of gases (99.7% He and 0.3% $N_2$),. The $1/e^2$ intensity radius of the laser focus spot was 28 μm, giving a Rayleigh length $Z_r$ of 3.1 mm. The Strehl ratio of the focus spot was 0.4–0.5. Thus, the peak focused laser intensity and the corresponding normalized vector potential, $a_0$, are approximately 2.2–3.1×10$^{18}$ W/cm$^2$ and 1.0–1.2, respectively. Two DRZ fluorescent screens were used in the detection system this time. The first one (area of 5×5 cm$^2$) was imaged into a 16-bit CCD camera (DU920P-BVF, Andor). It was placed before a 6-cm-long permanent dipole magnet to online diagnose the spatial profile and pointing angle of the electron beams. The second DRZ with the size of 30 (length) × 10 (width) cm$^2$, was imaged into an ICCD camera (DH334T-18U-63, Andor), it was placed after the magnet to online diagnose energy spectrum of the electron beams deflected by the effective magnetic field of ~$B_{eff}$ = 0.9 T. The distances from the first DRZ screen, the magnet, and the second DRZ screen to the gas jet were 72, 81, and 161 cm, respectively. We revised our MATLAB electron trajectory code used in Ref. [29] by involving the pointing angle of an electron beam incident on the magnet [30], and used this new code to calculate the electron energy spectrum more accurately. An integrating current transformer (ICT) and the top-view monitoring system were also used to measure the accelerated electron beam charge and observe the laser-plasma interaction, respectively. By varying the stagnation pressure of the gas jet from 2 atm to 5.1 atm, we can estimate the plasma density ($n_e$) to be 3.3–8.5×10$^{18}$ cm$^{-3}$ [29].

## 3. Results and discussions

Typical energy spectra of laser-driven accelerated electron beams from both pure He gas jet, and a mixture of 99.7 % He with 0.3% $N_2$ at a similar plasma density and laser power (7.5×10$^{18}$ cm$^{-3}$, 33 TW and 6.7×10$^{18}$ cm$^{-3}$, 31 TW) are shown in Fig. 1. It is shown that the peak energy of electron beam has been significantly enhanced from 145±5 MeV in pure helium gas jet to 275±5 MeV in the 0.3 % nitrogen-doped helium gas jet. Also, the maximum energy of the electron beam has been boosted to 500 MeV, that is 3.3 times higher than the beams from pure He case. The phenomenon of double-quasimonoenergetic energy peaks (Fig. 1(a)) due to sequential or multiple electron self-injections [31] in pure He plasma has disappeared in the gas mixture plasma. Figure 1(b) shows electron beam with only one peak without a dark current, in contrary to the nearly-continuous electron beams generated in previous ionization injection experiments where large amounts of dopant gas were mixed with the helium [22–24,32]. In addition, we can find that, even under the condition of lower laser power and plasma density, the beam charge in the mixed gas case (67.7 pC) are higher than that in the pure helium case (40.1 pC). We will continue to discuss the enhancement in the beam charge in the following contents.

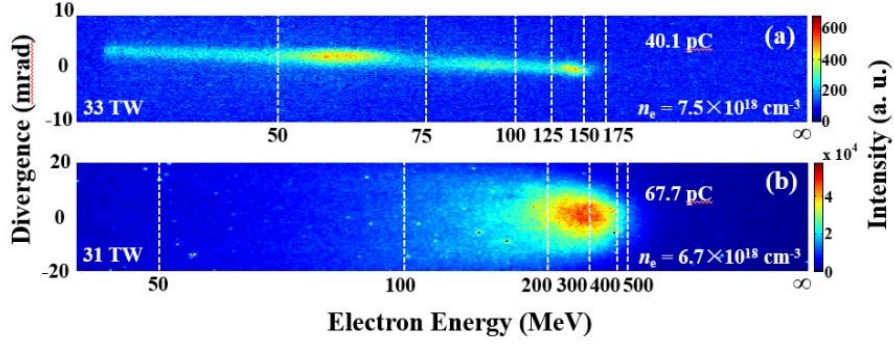

Fig. 1. A comparison between the observed raw energy spectra of electron bunches generated from (a) pure He gas at the density of $7.5\times10^{18}$ cm$^{-3}$ and (b) the mixing gas of 0.3% $N_2$ mixed with 99.7 % helium the plasma density of $6.7\times10^{18}$ cm$^{-3}$.

Figure 2 shows the electron energy spectra of six sequential shots obtained from the mixture of nitrogen and helium at the helium electron density of $5\times10^{18}$ cm$^{-3}$ and $6.7\times10^{18}$ cm$^{-3}$ and nitrogen plasma density $1.5\times10^{16}$ cm$^{-3}$ and $2.0\times10^{16}$ cm$^{-3}$. The position of energy labels in each panel is different due to the different pointing angle of each incident electron beam onto the magnet. The energy spectra demonstrate again the single electron bunch generation; the feature of no dark current. Our measurement statistics have shown that in the mixed gas jet target ~70% of the total shots have generated electron beams with higher than 200 MeV peak energy at the helium electron density range of $3.3\times10^{18}$ cm$^{-3}$ to $8.5\times10^{18}$ cm$^{-3}$ due to the ionization injection from the nitrogen; while there was no such high energy electron beam generation from the self-injection in pure helium gas jet at the same density range. This reveals the fact that it is easier to generate higher energy electrons using the ionization injection. From Figs. 2(a)–2(d), we can see the trend of increasing the electron beam peak energy with the laser power, such trend will be further proved in Fig. 4.

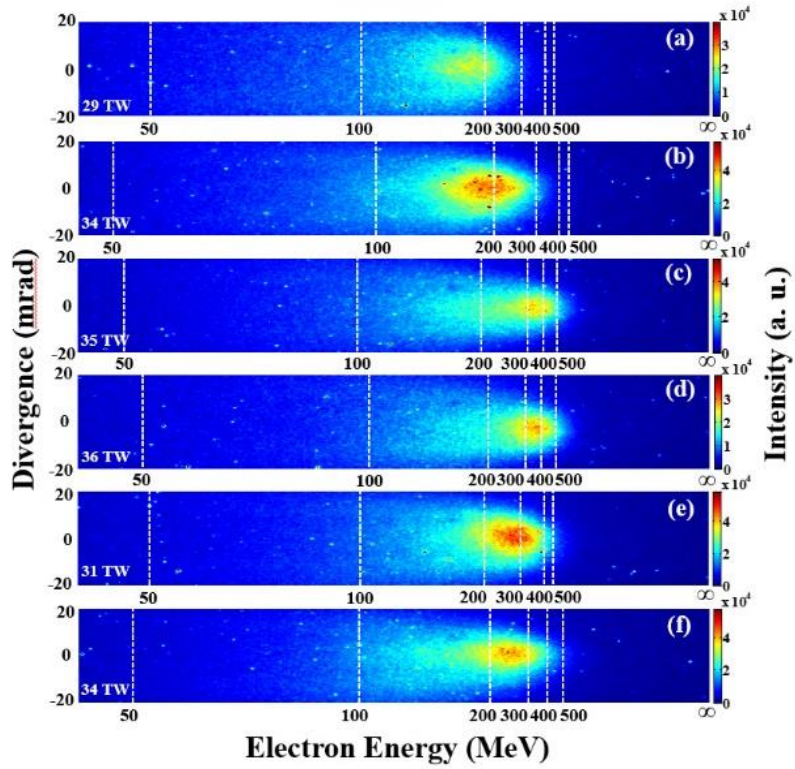

Fig. 2. Electron energy spectra of six successive shots for laser-driven gas jet of a 0.3 % nitrogen mixed with 99.7 % helium at the helium plasma density of (a)–(d) $5\times10^{18}$ cm$^{-3}$ and (e)–(f) $6.7\times10^{18}$ cm$^{-3}$; the laser power is written on each panel

The electron beam charge is plotted in Fig. 3 as a function of the helium plasma density in the range of 3.3–8.5×10$^{18}$ cm$^{-3}$ for the case of He mixed with 0.3% N$_2$ and compared to those for the case of pure He reported in our previous work of Ref. [29]. The charge here was measured by taking the average of all shots at the same gas pressure in the same experimental day. When the plasma density was higher than 8.5×10$^{18}$ cm$^{-3}$, we found the quality of electron beams degraded dramatically, so we stopped continuing to increase the gas pressure and focused on the above range of plasma density. Focusing on the density threshold for the pure He and its mixture with nitrogen, one can find that the addition of 0.3% N$_2$ into the helium has lowered the electron trapping threshold from 4.8×10$^{18}$ cm$^{-3}$ to 3.3×10$^{18}$ cm$^{-3}$, due to the ionization injection from the nitrogen. This means that electrons could be accelerated through a longer dephasing length to higher energy in the ionization injection scheme. In addition, what can be seen from Fig. 3 is an increase of the peak charge obtained with the addition of 0.3% N$_2$ into 99.7% He in the same range of plasma density. Our experiment and the experiment of Mo *et al.* [24] have a very good agreement on the enhancement of the electron beam charge and on lowering the density threshold for electron trapping due to the ionization injection.

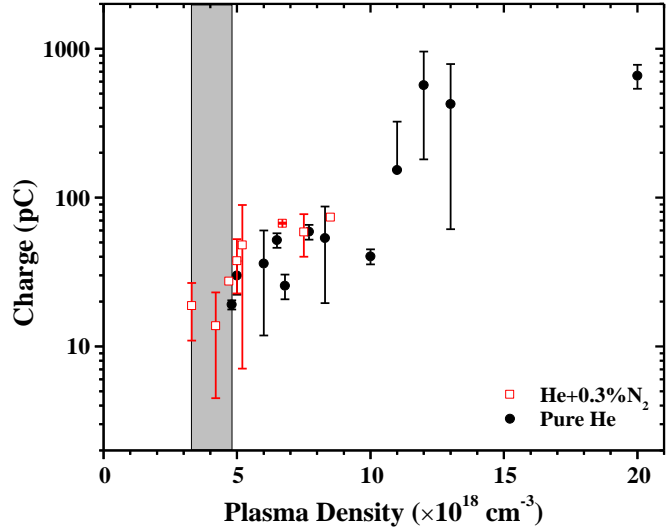

Fig. 3. Measured electron beam charge versus the plasma density for two different targets: pure He (black solid circles) [29] and 99.7 % He mixed with 0.3% $N_2$ (red empty squares). The greyish shaded region represents the region that is below the injection threshold of helium electron density of $4.8 \times 10^{18}$ cm$^{-3}$ for pure He to the injection threshold of $3.3 \times 10^{18}$ cm$^{-3}$ for the gas mixture.

Figure 4 plots the dependence of the electron beam average peak energy (empty circles) and maximum energy (solid triangles) on the laser power (Fig. 4(a)) and plasma density (Fig. 4(b)). The theoretically predicted maximum electron energy scaling (red lines) curves for a given plasma density of $5 \times 10^{18}$ cm$^{-3}$ and laser power of 30 TW are plotted in Fig. 4(a) and (b) as well, according to the relation [33]:

$$\Delta E(\text{GeV}) \cong 1.7 \left(\frac{P}{100 \text{ TW}}\right)^{\frac{1}{3}} \left(\frac{10^{18} \text{ cm}^{-3}}{n_e}\right)^{\frac{2}{3}} \left(\frac{0.8 \text{ µm}}{\lambda_0}\right)^{\frac{4}{3}}, \qquad (1)$$

where $P$ is the laser power, $n_e$ is the plasma density, and $\lambda_0$ is the central wavelength (800 nm) of laser. There is a clear trend shown in Fig. 4(a): as the laser power increases, the maximum and average peak energy of electron beams increases, until the energy show some sort of saturation at higher laser power. Also, the trend of higher electron energies at lower plasma densities is clear in Fig. 4(b). The theoretically predicted energies are mostly higher than those observed experimentally; this is because the theory assumes that the electrons are accelerated at the maximum value of the wakefield because the electrons in this case are injected at the wake bubble base. While, in the case of ionization injection the electrons are injected at the laser position in front of the bubble, so electrons can ride at different phases of the wakefield at which the field is not the maximum, thus the energy is lower.

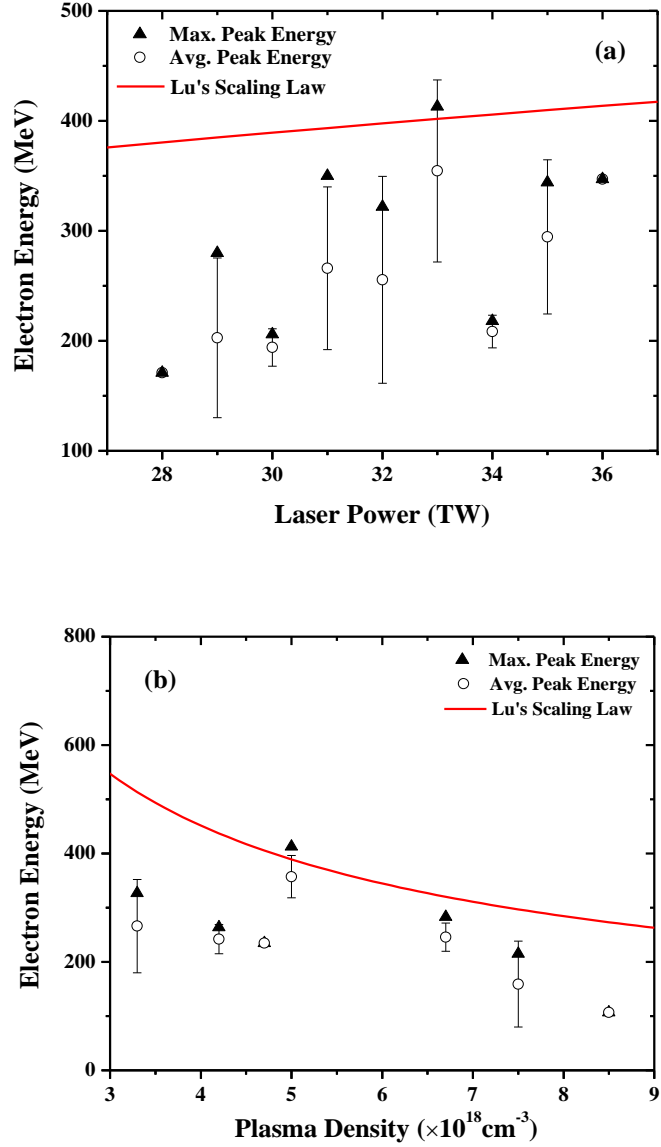

Fig. 4. The maximum peak energies (solid triangles) and average peak energies (empty circles) in electron energy spectra obtained from He mixed with 0.3% $N_2$ as a function of (a) the laser power and (b) the plasma density. The standard deviation of peak energies is considered as the error bar in both graphs. The data in (a) is measured at the plasma density of $5\times10^{18}$ cm$^{-3}$, and the data in (b) is measured at the laser power of 30 TW. The red lines in (a) and (b) represent the energies predicted by the empirical law of Eq. (1).

Recently, multidimensional simulations and theoretical results reported by Zeng

*et al.* [27] may explain some phenomena observed in our experiment. They proposed a self-truncated ionization injection LWFA scheme based on the use of ultralow traces of nitrogen (< 1%) in helium gas and unmatched laser spot size for the generation of electron beams of 1% level low-energy spread. The matched laser should satisfy the following condition, $(k_p\omega_0 = 2\sqrt{a_0})$. When using an initially unmatched laser pulse to shoot at the mixture gas target, the self-focusing of laser pulse leads to a strong wakefield evolution which subsequently breaks the ionization-induced injection condition and automatically truncate the injection process. As a result, the produced electron beam will have a narrow energy spread and no dark current in its energy spectra. In our experimental case of Fig. 1(b), $a_0$ = 1.1 and $k_p\omega_0$ = 13.66, thus we are using highly unmatched laser-plasma experimental condition.. The normalized laser power $P/P_c = (k_p\omega_0 a_0)^2/32$ is calculated to be 7.06, which is much greater than 1, which means that the self-focusing of laser pulse must occur and subsequently the ionization injection can occur and truncate. We can estimate from Ref. [27] the ionization injection length to be $z_{\text{cut}} = Z_r(\sqrt{2}\frac{P}{P_c} - 1)^{-1/2}$ = 1 mm. Such length is three times shorter than the length of plasma channel (~3 mm). That means electrons tunneling ionized from the *K*-shell of nitrogen atom were injected into the bubble and the injection process would be automatically truncated due to self-focusing of laser pulse in the first 1-mm length of plasma, then these electrons would be accelerated in the next 2-mm length. As a result, electron beams having high energies with low energy spread and no dark current were produced.

## 4. Conclusions

In conclusion, in the interaction of highly unmatched laser pulses with helium gas jet mixed with low ratio (0.3%) of nitrogen, we experimentally demonstrated overall enhanced laser wakefield acceleration of electron beams via ionization induced injection. Electron bunches with average energies of the order of 300 MeV have been produced using ~30 TW laser pulses. The use of ionization injection has led to increased electron charge, lower injection threshold plasma density, and much higher probability of generating dark-current free single electron bunches above 200 MeV as compared to self-injection in pure He under the same condition. An immediate possible mechanism to explain the results is the self-truncated ionization injection due to self-focusing which was recently proposed in Ref. [27]. It is foreseen that further optimization of such a scheme is expected to bring the electron beam energy-spread down to 1 %, making them suitable for driving ultra-compact free-electron lasers [28].

**Acknowledgments**

This work was supported by the National "973" Program of China (Grant No. 2013CBA01504) and the National Natural Science Foundation of China (Grants: 11121504, 11334013, 11175119, and 11374209).